\documentclass[12pt]{revtex4}  

\usepackage{graphicx}
\usepackage{amsmath} 
\usepackage{textcomp}
\usepackage{mathrsfs}
\usepackage{amsfonts}

\begin{document}
\newcommand{\vett}[1]{\mathbf{#1}}
\newcommand{\uvett}[1]{\hat{\vett{#1}}}
\newcommand{\beq}{\begin{equation}}
\newcommand{\eeq}{\end{equation}}
\newcommand{\barr}{\begin{eqnarray}}
\newcommand{\earr}{\end{eqnarray}}
\newcommand{\bseq}{\begin{subequations}}
\newcommand{\eseq}{\end{subequations}}

\title{Cylindrically Polarized Nondiffracting Optical Pulses}

\author{Marco Ornigotti$^{1,*}$, Claudio Conti$^{2,3}$ and Alexander Szameit$^1$}
\affiliation{$^1$Institute of Applied Physics, Friedrich-Schiller Universit\"at Jena, Max-Wien Platz 1, 07743 Jena, Germany}
\affiliation{$^2$ Institute for Complex Systems (ISC-CNR), Via dei Taurini 19, 00185, Rome, Italy}
\affiliation{$^3$  Department of Physics, University Sapienza, Piazzale Aldo Moro 5, 00185, Rome, Italy}
\email{*marco.ornigotti@uni-jena.de}


\begin{abstract}
We extend the concept of radially and azimuthally polarized optical beams to the polychromatic domain by introducing cylindrically polarized nondiffracting optical pulses. In particular, we discuss in detail the case of cylindrically polarized X-waves, both in the paraxial and nonparaxial regime. The explicit expressions for the electric and magnetic fields of cylindrically polarized X-waves is also reported.
\end{abstract}

\pacs{XXX}
\date{\today}
\maketitle

\section{Introduction}
Cylindrically polarized vector beams, namely solutions of Maxwell's equation with either radially or azimuthally distributed polarization across their intensity profile, have attracted a lot of interest in the past decade, thanks to their peculiar properties, such as the ability of producing a smaller focus \cite{ref1}, which opened new possibilities for spectroscopy \cite{ref2}, microscopy \cite{ref3}, particle manipulation \cite{ref4}, material processing \cite{ref6,ref7}, propagation of linear and nonlinear waves in crystals \cite{ref8,ref9,ref10,ref11}, quantum information \cite{ref13}, metrology \cite{ref14} and sensing \cite{ref15}. This vast range of applications motivated the development of different ways to efficiently generate such modes \cite{ref16,ref17,ref18,ref19,ref20}. A detailed theoretical discussion on the properties and applications of these beams in the paraxial case can be found in \cite{ragazzina}. Inspired by all this, many groups have tried to extend the properties and structure of these beams to the nonparaxial case, by investigating strongly focused fields \cite{ref23}, complex dipole sources \cite{ref24}, elegant \cite{ref25} and decentered \cite{ref26} Laguerre-Gauss beams, and vector Bessel beams \cite{ref27}. Recently, cylindrically polarized Bessel-Gauss beams have also been theoretically proposed \cite{ref29}. Very recently, moreover, the concept of radial and azimuthal polarization has been investigated experimentally also for optical pulses, leading to some interesting results in material processing \cite{ref31}, focusing of femtosecond pulses \cite{ref32} and electron-photon interactions \cite{ref33}. A fully theoretical description of cylindrically polarized optical pulses has, however, not yet been proposed.

It is  the aim of this work to fill this gap. As suggested in \cite{ragazzina}, to properly construct cylindrically polarized modes, one needs states of the electromagnetic field with one unit of orbital angular momentum (OAM) and one unit of spin angular momentum (SAM), combined in such a way that the \emph{total} angular momentum of the field will remain zero. In the paraxial case, this is achieved by using linearly polarized Hermite-Gaussian beams \cite{ragazzina}, while in the nonparaxial domain, suitable superpositions of Bessel beams are used to mimick the structure of the correspondent paraxial modes \cite{mioOptExpr}. Here, we use the latter approach to extend the definition of cylindrically polarized modes to the domain of optical pulses. Polychromatic superpositions of Bessel beams, commonly called X-waves, in fact, have been studied quite extensively in the past years \cite{localizedWaves}, and very recently, OAM-carrying X-waves have been proposed \cite{nostroPRL}. Firstly introduced in acoustics \cite{ref35, ref36}, X-waves have provided many interesting results in many different areas physics, like nonlinear optics \cite{ref37,ref38}, condensed matter \cite{ref39}, quantum optics \cite{ref40}, integrated optics \cite{ref41,ref41bis} and optical communications \cite{ref42}. We believe that the interplay between the nondiffracting character typical of X-waves and the properties of cylindrically polarized beams can open new possibilities for fundamental and applied research.

This work is organized as follows: in Sect. 2, we briefly review how to obtain scalar and vector X-waves carrying OAM starting from Bessel beams. In Sect. 3, we introduce radially and azimuthally polarized optical pulses, and use these results in Sect. 4 to discuss the special case of cylindrically polarized X-waves. Finally, conclusions are drawn in Sect. 5.

\section{X Waves Carrying OAM}
In this section, we briefly discuss how to construct nondiffracting optical pulses carrying OAM. To do that, we essentially follow the method presented in Ref. \cite{nostroPRA}. Let us consider a general solution $\psi(\vett{r}; k)$ of the scalar Helmholtz equation, i.e., 
\beq\label{eqZero}
(\nabla^2+k^2)\psi(\vett{r}; k)=0,
\eeq
where $k=\omega/c$ is the vacuum wave vector. From this solution, it is possible to construct an exact solution of the scalar wave equation $ \left(\nabla^2-\frac{1}{c^2}\partial_t^2\right)\phi(\vett{r},t)=0$,
  in the following way:
\beq\label{eqUno}
\phi(\vett{r},t)=\int_0^{\infty}\,d\,k\,f(k)\,e^{-ic k t}\,\psi(\vett{r}; k),
\eeq
where $f(k)$ is an arbitrary spectral function. If for $\psi(\vett{r}; k)$ we choose Bessel beams i.e., nondiffracting solutions of the Helmholtz equation, then, according to Ref. \cite{durnin}, the solution to Eq. \eqref{eqZero} can be written as
\beq
\psi_m^B(\vett{r}; k)=\text{J}_m\left(k\sin\vartheta_0 R \right)e^{im\theta}e^{ik z\cos\vartheta_0},
\eeq
where $\theta=\arctan(y/x)$, $R=\sqrt{x^2+y^2}$ (being $x$,$y$ the coordinates in the transverse plane orthogonal to the propagation direction $z$),  $J_m(x)$ is the Bessel function of the first kind of order $m$ \cite{nist} and $\vartheta_0$ is the Bessel cone angle, i.e., the beam's characteristic parameter. X-waves are then defined, according to Eq. \eqref{eqUno}, as polychromatic superpositions of Bessel beams \cite{localizedWaves} as follows:
\barr\label{eq1}
\phi_m(\vett{r},t)=\int_0^{\infty}\,d\,k\,f(k)\,J_m(k\sin\vartheta_0 R)\,e^{i(k\zeta+m\theta)},
\earr
 where $\zeta=z\cos\vartheta_0-ct$ is the co-moving coordinate attached to the nondiffracting pulse itself and the subscript $m$ has been added to emphasize the fact that the above scalar field carries $m$ units of OAM. For the case $m\neq 0$, the above equation describes a scalar nondiffracting wave carrying $m$ units of OAM \cite{nostroPRL}. 
\subsection{Generalization to the Vector Case}

The results presented above are only valid in the scalar case. To obtain vectorial X-waves that are exact solutions to Maxwell's equation, we can use the following simple vectorialization procedure, based on the Hertz potential method \cite{stratton}. If we introduce the monochromatic Hertz potential $\textbf{P}(\vett{r},t; k)=\psi(\vett{r}; k)\exp{(-ick t)}\uvett{f}$ (where $\uvett{f}$ is an arbitrary unit vector), in fact, the correspondent monochromatic electric and magnetic fields $\mathcal{E}(\vett{r},t; k)$ and $\mathcal{B}(\vett{r},t; k)$ are defined via the following relations \cite{mioHertz}: 
\bseq\label{eqHertz}
\begin{align}
\mathcal{E}(\vett{r},t; k) &= \nabla\times\nabla\times\textbf{P}(\vett{r},t; k),\\
\mathcal{B}(\vett{r},t; k) &= \frac{1}{c^2}\frac{\partial}{\partial t}\left[\nabla\times\textbf{P}(\vett{r},t; k)\right].
\end{align}
\eseq
It is then easy to verify that the above fields are exact solutions of Maxwell's equations. Once these fields are known, the electric and magnetic field corresponding to the nondiffracting optical field given by Eq. \eqref{eqUno} are given by
\bseq\label{polyFields}
\begin{align}
\vett{E}(\vett{r},t)&=\int_0^{\infty}\,d\,k\,f(k)\,\mathcal{E}(\vett{r},t; k),\\
\vett{B}(\vett{r},t)&=\int_0^{\infty}\,d\, k\,f(k)\,\mathcal{B}(\vett{r},t; k).
\end{align}
\eseq
Note that, alternatively, one could directly define the polychromatic Hertz potential $\boldsymbol\Pi(\vett{r},t)=\phi_m(\vett{r},t)\uvett{f}$ and use Eqs. \eqref{eqHertz} with $\boldsymbol\Pi$ instead of $\vett{P}$ to directly generate the polychromatic electric and magnetic fields $\vett{E}(\vett{r},t)$ and $\vett{B}(\vett{r},t)$. 
\subsection{Fundamental X-waves with OAM}

As an explicit example of optical pulse carrying OAM, we consider the so-called fundamental X-waves \cite{localizedWaves}, which possess the following exponentially decaying spectrum \cite{refClaudio}:
\beq\label{eqSpectrum}
f(k)=k^ne^{-\alpha k},
\eeq
where $\alpha$ is a constant, which essentially accounts for the spectrum width, and $n\in\mathrm{N}$. The spectrum in Eq. \eqref{eqSpectrum} can be obtained from the generating function $\exp{(-\alpha k)}$ by subsequent differentiations with respect to $\alpha$, i.e., $k^n\exp{(-\alpha k)}=(-1)^n\partial_{\alpha}^n\left[\exp{(-\alpha k)}\right]$. For the sake of simplicity, therefore, here we will only consider the case $n=0$. The higher order fields corresponding to $n\neq 0$ can be then obtained by our fundamental solution by simple differentiation with respect to $\alpha$. Substituting \eqref{eqSpectrum} with $n=0$ into Eq. \eqref{eq1} and using the formula 6.621.1 of Ref. \cite{gradstein}, gives then the following result:
\beq\label{xOAM}
\phi_m(\vett{r},t)=\frac{1}{2^m}\left[\frac{\xi^m}{(\alpha-i\zeta)}\right]e^{im\theta}\, _2F_1\left(\frac{m+1}{2},\frac{m+2}{2};m+1;-\xi^2\right),
\eeq
where $\, _2F_1(a,b;c;x)$ is the Gauss' hypergeometric function \cite{nist} and $\xi=R\sin\vartheta_0/(\alpha-i\zeta)$. The higher-order solutions, namely the one corresponding to  $n\neq 0$ in Eq. \eqref{eqSpectrum}, are then obtained by taking the $n$-th derivative of Eq. \eqref{xOAM} with respect to $(\alpha-i\zeta)$. 

If we introduce, in analogy with the paraxial beams \cite{siegman}, the beam waist $w(\zeta)$ as
\beq
w(\zeta)=w_0\left(1-i\frac{\zeta}{\alpha}\right),
\eeq
where $w_0\equiv w(\zeta=0)=\alpha/\sin\vartheta_0$, and introduce the normalized radial variable $\rho\equiv\rho(\zeta)= R/w(\zeta)$, Eq. \eqref{xOAM} can be then rewritten in the following form:
\beq\label{xOAM2}
\phi_m(\rho,\theta,\zeta)=\frac{\rho^{m+1}}{2^m\sin\vartheta_0}e^{im\theta}\, _2F_1\left(\frac{m+1}{2},\frac{m+2}{2};m+1;-\rho^2\right).
\eeq
Equation \eqref{xOAM2} represents a scalar nondiffracting optical pulse  carrying $m$ units of OAM. The vector electric and magnetic fields for fundamental X-waves with OAM can be calculated with Eqs. \eqref{eqHertz} by choosing $\vett{P}(\vett{r},t)=\psi^B_m(\vett{r}; k)\exp{(-ickt)}\uvett{f}$ and using the spectrum \eqref{eqSpectrum} with $n=0$. Their explicit expression can be found in Ref. \cite{nostroPRA}.
\section{Cylindrically Polarized Nondiffracting Optical Pulses: General Aspects}

Cylindrically polarized beams are a special subset of vector beams with non-uniform polarization. They can be thought of being defined as linear combinations of basis vector of a four dimensional vector space $\mathcal{H}=\mathcal{H}_M\otimes\mathcal{H}_P$, defined as the cartesian product of two 2-dimensional vector spaces, namely  the space $\mathcal{H}_M$, spanned by the two mode functions $\{\Psi_{10}(\vett{r}),\Psi_{01}(\vett{r})\}$ and the space $\mathcal{H}_P$, spanned by the two polarization vectors $\{\uvett{x},\uvett{y}\}$ \cite{ragazzina}. Therefore, $\mathcal{H}=\{\Psi_{10}(\vett{r})\uvett{x},\Psi_{10}(\vett{r})\uvett{y},\Psi_{01}(\vett{r})\uvett{x},\Psi_{01}(\vett{r})\uvett{y}\}$. The explicit form of the mode functions $\Psi_{10}(\vett{r})$ and $\Psi_{01}(\vett{r})$, however, depends on whether we consider paraxial or nonparaxial beams. For the paraxial case, the mode functions are simply the Hermite-Gauss beams $HG_{nm}(\vett{r})$ of order $N=n+m=1$ \cite{siegman}, namely $\{\Psi_{10}(\vett{r}),\Psi_{01}(\vett{r})\}\equiv\{HG_{10}(\vett{r}),HG_{01}(\vett{r})\}$ \cite{ragazzina}. In the nonparaxial case, instead, the mode functions $\Psi_{10}(\vett{r})$ and $\Psi_{01}(\vett{r})$ can be written in terms of superpositions of Bessel beams with $m=\pm 1$, as follows \cite{mioOptExpr}:
\bseq\label{NPmodes}
\begin{align}
\Psi_{10}(\vett{r}; k) &=\frac{1}{\sqrt{2}}\left[\psi_1^B(\vett{r}; k)+\psi_{-1}^B(\vett{r}; k)\right],\\
\Psi_{01}(\vett{r}; k) &=\frac{-i}{\sqrt{2}}\left[\psi_1^B(\vett{r}; k)-\psi_{-1}^B(\vett{r}; k)\right].
\end{align}
\eseq
Linear combinations of the four basis vectors spanning $\mathcal{H}$ give then rise to radially (R) and azimuthally (A) polarized fields, as follows:
\bseq\label{modes}
\begin{align}
\uvett{u}_R^{\pm}(\vett{r}; k) &=\frac{1}{\sqrt{2}}\left[\pm \Psi_{10}(\vett{r})\uvett{x}+\Psi_{01}(\vett{r})\uvett{y}\right],\\
\uvett{u}_A^{\pm}(\vett{r}; k) &=\frac{1}{\sqrt{2}}\left[\mp \Psi_{01}(\vett{r})\uvett{x}+\Psi_{10}(\vett{r})\uvett{y}\right],
\end{align}
\eseq
where the $\pm$ determine the co-rotating and counter-rotating solutions, respectively \cite{ragazzina}. The results presented above are valid for monochromatic fields only. To generalize the mode functions $\Psi_{10}(\vett{r})$ and $\Psi_{01}(\vett{r})$ to the polychromatic case, we first substitute Eqs. \eqref{NPmodes} into Eq. \eqref{eqUno}, thus obtaining
\bseq\label{HGbessel}
\begin{align}
\Phi_{10}(\vett{r},t) &=\int_0^{\infty}\,d\,kf\;(k)e^{-ic k t}\Psi_{10}(\vett{r}; k),\\
\Phi_{01}(\vett{r},t) &=\int_0^{\infty}\,d\,kf\;(k)e^{-ickt}\Psi_{01}(\vett{r}; k).
\end{align}
\eseq

Next, we use the scalar pulses defined above into Eqs. \eqref{modes}, to obtain the cylindrically polarized vector pulses:
\bseq\label{modesPulses}
\begin{align}
\uvett{U}_R^{\pm}(\vett{r},t) &=\int_0^{\infty}\,d\,k\,f(k)\,e^{-ickt}\uvett{u}_R^{\pm}(\vett{r}; k)=\frac{1}{\sqrt{2}}\left[\pm \Phi_{10}(\vett{r},t)\uvett{x}+\Phi_{01}(\vett{r},t)\uvett{y}\right],\\
\uvett{U}_A^{\pm}(\vett{r},t) &=\int_0^{\infty}\,d\,k\,f(k)\,e^{-ickt}\uvett{u}_A^{\pm}(\vett{r}; k)=\frac{1}{\sqrt{2}}\left[\mp \Phi_{01}(\vett{r},t)\uvett{x}+\Phi_{10}(\vett{r},t)\uvett{y}\right].
\end{align}
\eseq
The equations above are the first result of our work. They are, essentially, a generalization of the correspondent radially and azimuthally polarized vector modes (introduced in Refs. \cite{ragazzina} and \cite{mioOptExpr}) to the case of optical pulses. By choosing the form of $\Psi_{10}(\vett{r})$ and $\Psi_{01}(\vett{r})$ within the paraxial or nonparaxial regime, Eqs. \eqref{modesPulses} define paraxial and nonparaxial cylindrically polarized pulses, respectively. As discussed in detail in Ref. \cite{mioOptExpr}, however, in the nonparaxial regime the vector fields in Eqs. \eqref{modesPulses} do not constitute an exact solution to Maxwell's equations, as they fail to satisfy the transversality condition $\nabla\cdot\vett{U}_{R,A}^{\pm}=0$. To overcome this problem, we can employ the vectorialization procedure described in Sect. 3. If we choose $\vett{P}(\vett{r},t)=\vett{u}_{R,A}^{\pm}(\vett{r}; k)\exp{(-ickt)}$ as Hertz potentials, then the electric and magnetic fields of a cylindrically polarized optical pulse can be written, in the general case, as follows:
\bseq\label{cylFields}
\begin{align}
\vett{E}_{R,A}^{\pm}(\vett{r},t)&=\int_0^{\infty}\,d\,k\,f(k)\,\mathcal{E}_{R,A}^{\pm}(\vett{r},t; k),\\
\vett{B}_{R,A}^{\pm}(\vett{r},t)&=\int_0^{\infty}\,d\, k\,f(k)\,\mathcal{B}_{R,A}^{\pm}(\vett{r},t; k),
\end{align}
\eseq
where, according to Eqs. \eqref{eqHertz}, $\mathcal{E}(\vett{r},t; k) = \nabla\times\nabla\times\textbf{P}(\vett{r},t; k)$ and $c^2\mathcal{B}(\vett{r},t; k) = \frac{\partial}{\partial t}\left[\nabla\times\textbf{P}(\vett{r},t; k)\right]$. The set of equations above represents the electric and magnetic fields of a general cylindrically polarized, nondiffracing optical pulse, and it is the second result of our paper. Equations \eqref{cylFields} describe a whole class of cylindrically polarized pulses, whose properties are defined by the choice of the spectrum $f(k)$.
\section{Cylindrically Polarized X-Waves}

\begin{figure}[!t]
\begin{center}
\includegraphics[width=\textwidth]{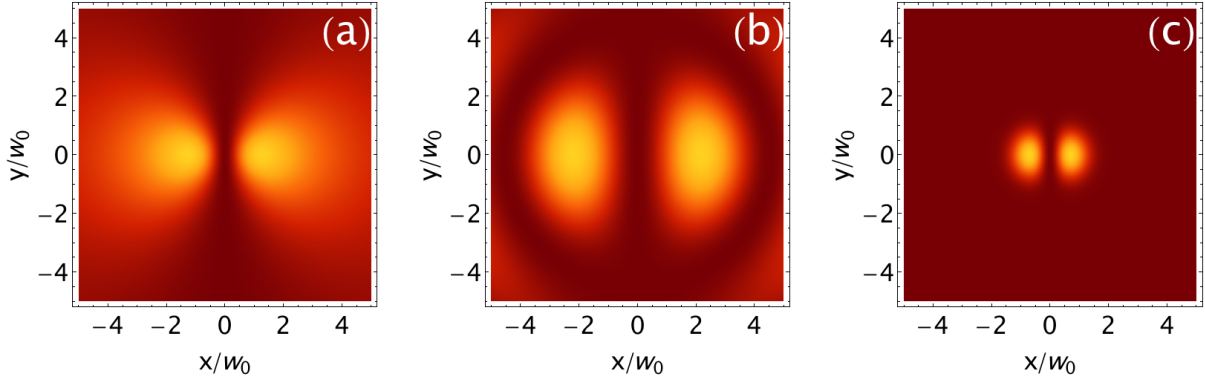}
\caption{Transverse cross section at $\zeta=0$ (corresponding to $z=0=t$) of (a) the scalar pulse $\Phi_{10}(\vett{r},t)$,  (b) the monochromatic scalar function 
$\Psi_{10}(\vett{r}; k)$ close to the propagation axis, and (c)  the monochromatic paraxial Hermite-Gauss beam $HG_{10}(x,y,0)$. A direct comparison between panels (a) (b) and (c)  shows that the symmetry possessed by the paraxial Hermite-Gaussian mode (c) is maintained also in the nonparaxial case (b), if the correspondent mode is built according to the first of Eqs. \eqref{NPmodes}. Moreover, the same symmetry is also preserved in the polychromatic regime for the scalar pulse $\Phi_{10}$, as panel (a) shows. For all panels, the transverse coordinates $\{x,y\}$ have been normalized to the correspondent beam waist $w_0$, which for panel (a) is given by $w_0=\alpha/\sin\vartheta_0$.}
\label{figure1}
\end{center}
\end{figure}

\begin{figure}[!t]
\begin{center}
\includegraphics[width=\textwidth]{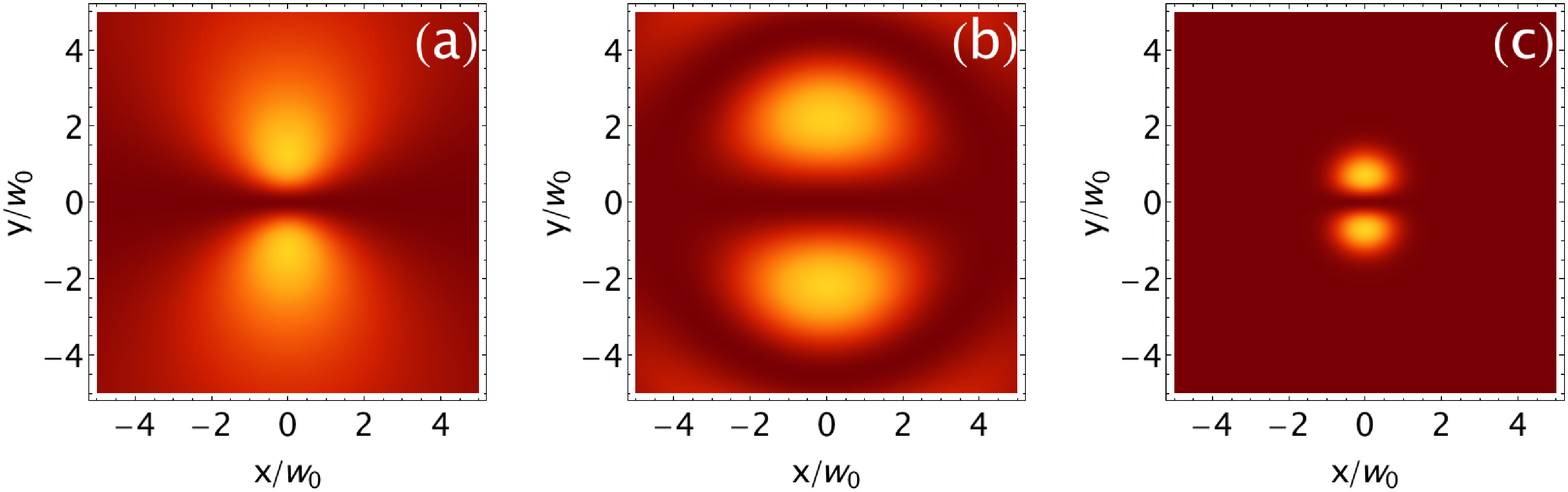}
\caption{Transverse cross section at $\zeta=0$ (corresponding to $z=0=t$) of (a) the scalar pulse $\Phi_{01}(\vett{r},t)$,  (b) the monochromatic scalar function 
$\Psi_{01}(\vett{r}; k)$ close to the propagation axis, and (c) the monochromatic paraxial Hermite-Gauss beam $HG_{01}(x,y,0)$. A direct comparison between panels (a) (b) and (c) shows that the symmetry possessed by the paraxial Hermite-Gaussian mode (c) is maintained also in the nonparaxial case (b), if the correspondent mode is built according to the second of Eqs. \eqref{NPmodes}. Moreover, the same symmetry is also preserved in the polychromatic regime for the scalar pulse $\Phi_{01}$, as panel (a) shows. For all panels, the transverse coordinates $\{x,y\}$ have been normalized to the correspondent beam waist $w_0$, which for panel (a) is given by $w_0=\alpha/\sin\vartheta_0$.}
\label{figure2}
\end{center}
\end{figure}

We now specify our analysis to the case of fundamental X-waves. In this case, Eqs. \eqref{HGbessel} have the following explicit form:
\bseq\label{HGpulses}
\begin{align}
\Phi_{10}(\rho,\theta,\zeta) &=\frac{\rho\cos\theta}{\sqrt{2}w(\zeta)}\, _2F_1\left(1,\frac{3}{2};2;-\xi^2\right),\\
\Phi_{01}(\rho,\theta,\zeta) &=\frac{\rho\sin\theta}{\sqrt{2}w(\zeta)}\, _2F_1\left(1,\frac{3}{2};2;-\xi^2\right).
\end{align}
\eseq
A comparison between these two functions and the correspondent monochromatic paraxial (Hermite-Gauss) and nonparaxial (Bessel) modes is reported in Figs. \ref{figure1} and \ref{figure2}. As it can be seen, the two pulses $\Phi_{10}$ and $\Phi_{01}$ retain the same symmetry of their monochromatic counterparts, and can be then taken as a good candidate to realize cylindrically polarized modes. Substituting these expressions into Eqs. \eqref{modesPulses} leads to the following result:
\bseq
\begin{align}
\vett{U}_R^{\pm}(\rho,\theta,\zeta) &=\frac{\rho}{2w(\zeta)}\, _2F_1\left(1,\frac{3}{2};2;-\xi^2\right)\left(\pm\cos\theta\uvett{x}+\sin\theta\uvett{y}\right),\\
\vett{U}_A^{\pm}(\rho,\theta,\zeta) &=\frac{\rho}{2w(\zeta)}\, _2F_1\left(1,\frac{3}{2};2;-\xi^2\right)\left(\mp\sin\theta\uvett{x}+\cos\theta\uvett{y}\right).
\end{align}
\eseq

The components of the vector electric field generated by $\vett{U}_{R,A}^{\pm}(\vett{r},t)$ can be then calculated by using the first of Eqs. \eqref{cylFields}, whose expression, in the co-moving cylindrical coordinates $\{\rho(\zeta),\theta,\zeta\}$, is given as follows:

\bseq\label{EradP}
\begin{align}
E_{R+}^r(\vett{R},\zeta) &= \frac{3\rho\cot^2\vartheta_0}{w^2(\zeta)}\, _2F_1\left(2,\frac{5}{2};2;-\rho^2\right),\\
E_{R+}^{\theta}(\vett{R},\zeta) &= 0,\\
E_{R+}^z(\vett{R},\zeta)&=\frac{2i\cot\vartheta_0}{\sin\vartheta_0w^3(\zeta)}\, _2F_1\left(\frac{3}{2},2;1;-\rho^2\right)\label{EradP3},
\end{align}
\eseq
for the co-rotating radially polarized field,
\bseq\label{EradM}
\begin{align}
E_{R-}^r(\vett{R},\zeta) &=\frac{\cos(2\theta)}{\rho^2w^3(\zeta)}\Big[2\sin\vartheta_0\, _2F_1\left(1,\frac{3}{2};1;-\xi^2\right)-2\, _2F_1\left(\frac{1}{2},1;1;-\rho^2\right)\nonumber\\
&- 3\rho^2\cos^2\vartheta_0\, _2F_1\left(2,\frac{5}{2};1;-\xi^2\right)\Big],\\
E_{R-}^{\theta}(\vett{R},\zeta) &= \frac{\sin(2\theta)}{\rho^2w^3(\zeta)}\Big[2\sin\vartheta_0\, _2F_1\left(1,\frac{3}{2};1;-\rho^2\right)-2\, _2F_1\left(\frac{1}{2},1;1;-\rho^2\right)\nonumber\\
&+ 3\rho^2\, _2F_1\left(2,\frac{5}{2};1;-\rho^2\right)\Big],\\
E_{R-}^z(\vett{R},\zeta)&=\frac{2i\cot\vartheta_0\cos(2\theta)}{\sin\vartheta_0w^3(\zeta)}\, _2F_1\left(\frac{3}{2},2;1;-\rho^2\right)\label{EradM3},
\end{align}
\eseq
for the counter-rotating radially polarized field,
\bseq\label{EazP}
\begin{align}
E_{A+}^r(\vett{R},\zeta) &=0, \\
E_{A+}^{\theta}(\vett{R},\zeta) &= \frac{3\rho}{w^2(\zeta)\sin^2\vartheta_0}\, _2F_1\left(2,\frac{5}{2};2;-\rho^2\right),\\
E_{A+}^z(\vett{R},\zeta) &= 0\label{EazP3},
\end{align}
\eseq
for the co-rotating azimuthally polarized field, and
\bseq\label{EazM}
\begin{align}
E_{A-}^r(\vett{R},\zeta)&=\frac{2\sin(2\theta)}{\rho^2w^3(\zeta)}\Big[-\sin\vartheta_0\, _2F_1\left(1,\frac{3}{2};1;-\rho^2\right)+\, _2F_1\left(\frac{1}{2},1;1;-\rho^2\right)\nonumber\\
&+ \frac{3}{2}\rho^2\cos^2\vartheta_0\, _2F_1\left(2,\frac{5}{2};1;-\rho^2\right)\Big],\\
E_{A-}^{\theta}(\vett{R},\zeta) &= \frac{2\cos(2\theta)}{\rho^2w^3(\zeta)}\Big[\sin\vartheta_0\, _2F_1\left(1,\frac{3}{2};1;-\rho^2\right)-\, _2F_1\left(\frac{1}{2},1;1;-\rho^2\right)\nonumber\\
&+ \frac{3}{2}\rho^2\, _2F_1\left(2,\frac{5}{2};1;-\rho^2\right)\Big],\\
E_{A-}^z(\vett{R},\zeta)&= -\frac{3i\rho^2\cos\vartheta_0\sin(2\theta)}{w(\zeta)}\, _2F_1\left(\frac{5}{2},3;3;-\rho^2\right)\label{EazM3},
\end{align}
\eseq
for the counter-rotating azimuthally polarized field. Here, $\vett{R}=\{\rho(\zeta)\uvett{r},\theta\boldsymbol\theta\}$, and $\zeta=z\cos\vartheta_0-ct$. In a similar manner, using the second of Eqs. \eqref{cylFields}, the components of the magnetic field of the cylindrically polarized fundamental X-waves are give as follows:
\begin{figure}[!t]
\begin{center}
\includegraphics[width=\textwidth]{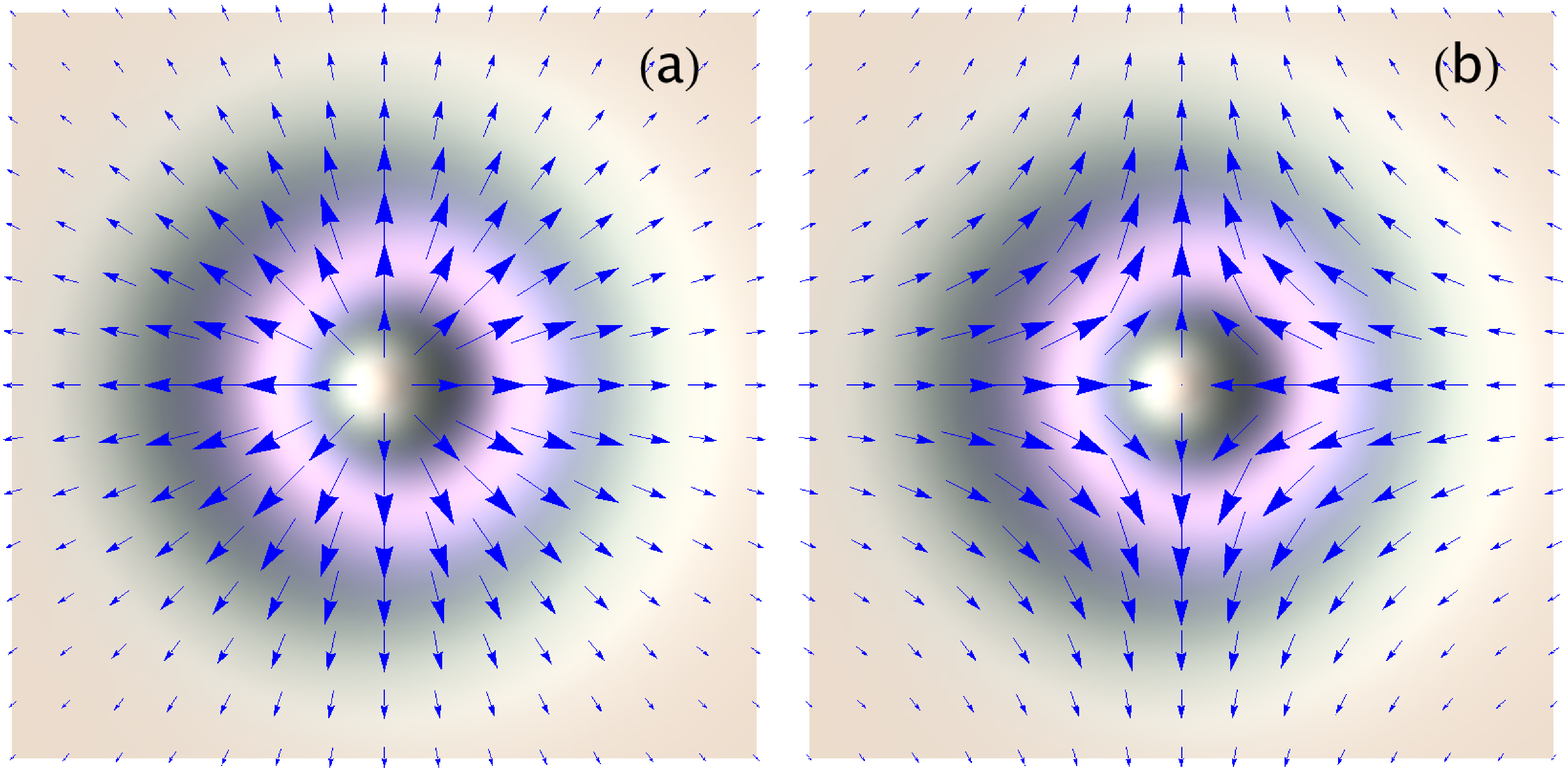}
\caption{Polarization pattern (arrows) and transverse intensity distribution $I(x,y)=|E_x(x,y,0)|^2+|E_y(x,y,0)|^2$ calculated at $\zeta=0$ (underlying doughnut) of the electric field of  (a) a co-rotating radially polarized X-wave, as defined in Eqs. \eqref{EradP}, and (b) a counter-rotating radially polarized X-wave, as defined in Eqs. \eqref{EradM}. The axes of both graphs span the interval $[-1.5,1.5]$ in units of the equivalent beam waist $w_0=\alpha/\sin\vartheta_0$.}
\label{figure3}
\end{center}
\end{figure}
\bseq
\begin{align}
cB_{R+}^r(\vett{R},\zeta) &= 0, \\
cB_{R+}^{\theta}(\vett{R},\zeta)&= \frac{3\rho\cot^2\vartheta_0}{w^2(\zeta)}\, _2F_1\left(2,\frac{5}{2};2;-\rho^2\right),\\
cB_{R+}^z(\vett{R},\zeta)&= 0 ,
\end{align}
\eseq
for the co-rotating radially polarized field,
\bseq
\begin{align}
cB_{R-}^r(\vett{R},\zeta) &= -\frac{3\rho\cot\vartheta_0\sin(2\theta)}{\sin\vartheta_0w^2(\zeta)}\, _2F_1\left(2,\frac{5}{2};2;-\rho^2\right) , \\
cB_{R-}^{\theta}(\vett{R},\zeta) &= - \frac{3\rho\cot\vartheta_0\cos(2\theta)}{\sin\vartheta_0w^2(\zeta)}\, _2F_1\left(2,\frac{5}{2};2;-\rho^2\right)  ,\\
cB_{R-}^z(\vett{R},\zeta)&= \frac{3i\rho^2\sin(2\theta)}{w(\zeta)}\, _2F_1\left(\frac{5}{2},3;3;-\rho^2\right) ,
\end{align}
\eseq
for the counter-rotating radially polarized field,
\bseq
\begin{align}
cB_{A+}^r(\vett{R},\zeta) &= - \frac{3\rho\cot\vartheta_0}{\sin\vartheta_0w^2(\zeta)}\, _2F_1\left(2,\frac{5}{2};2;-\rho^2\right), \\
cB_{A+}^{\theta}(\vett{R},\zeta) &= 0 ,\\
cB_{A+}^z(\vett{R},\zeta) &=- \frac{2i}{\sin^2\vartheta_0 w^3(\zeta)}\, _2F_1\left(\frac{3}{2},2;1;-\rho^2\right) ,
\end{align}
\eseq
for the co-rotating azimuthally polarized field, and
\bseq
\begin{align}
cB_{A-}^r(\vett{R},\zeta) &= -\frac{3\rho\cot\vartheta_0\cos(2\theta)}{\sin\vartheta_0 w^2(\zeta)}\, _2F_1\left(2,\frac{5}{2};2;-\rho^2\right), \\
cB_{A-}^{\theta}(\vett{R},\zeta) &= --\frac{3\rho\cot\vartheta_0\sin(2\theta)}{\sin\vartheta_0 w^2(\zeta)}\, _2F_1\left(2,\frac{5}{2};2;-\rho^2\right) ,\\
cB_{A-}^z(\vett{R},\zeta) &= \frac{3\rho^2}{w(\zeta)}\, _2F_1\left(\frac{5}{2},3;3;-\rho^2\right) ,
\end{align}
\eseq
for the counter-rotating azimuthally polarized field.

\begin{figure}[!t]
\begin{center}
\includegraphics[width=\textwidth]{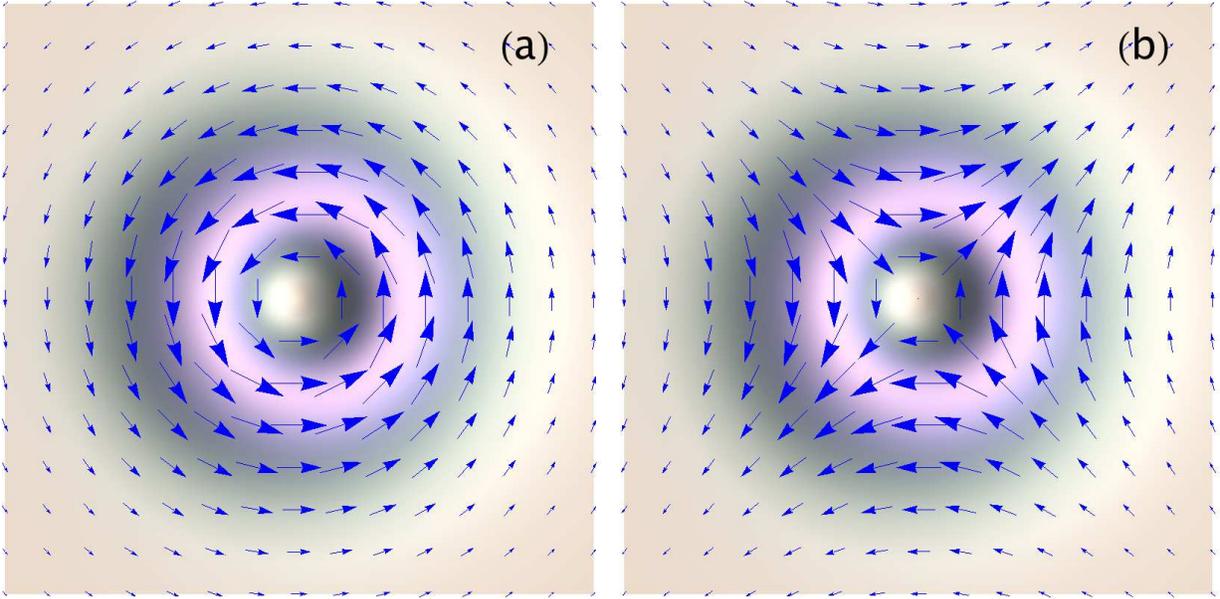}
\caption{Polarization pattern (arrows) and transverse intensity distribution $I(x,y)=|E_x(x,y,0)|^2+|E_y(x,y,0)|^2$ calculated at $\zeta=0$ (underlying doughnut) of the electric field of  (a) a co-rotating azimuthally polarized X-wave, as defined in Eqs. \eqref{EazP}, and (b) a counter-rotating azimuthally polarized X-wave, as defined in Eqs. \eqref{EazM}. The axes of both graphs span the interval $[-1.5,1.5]$ in units of the equivalent beam waist $w_0=\alpha/\sin\vartheta_0$.}
\label{figure4}
\end{center}
\end{figure}

\begin{figure}[!t]
\begin{center}
\includegraphics[width=0.8\textwidth]{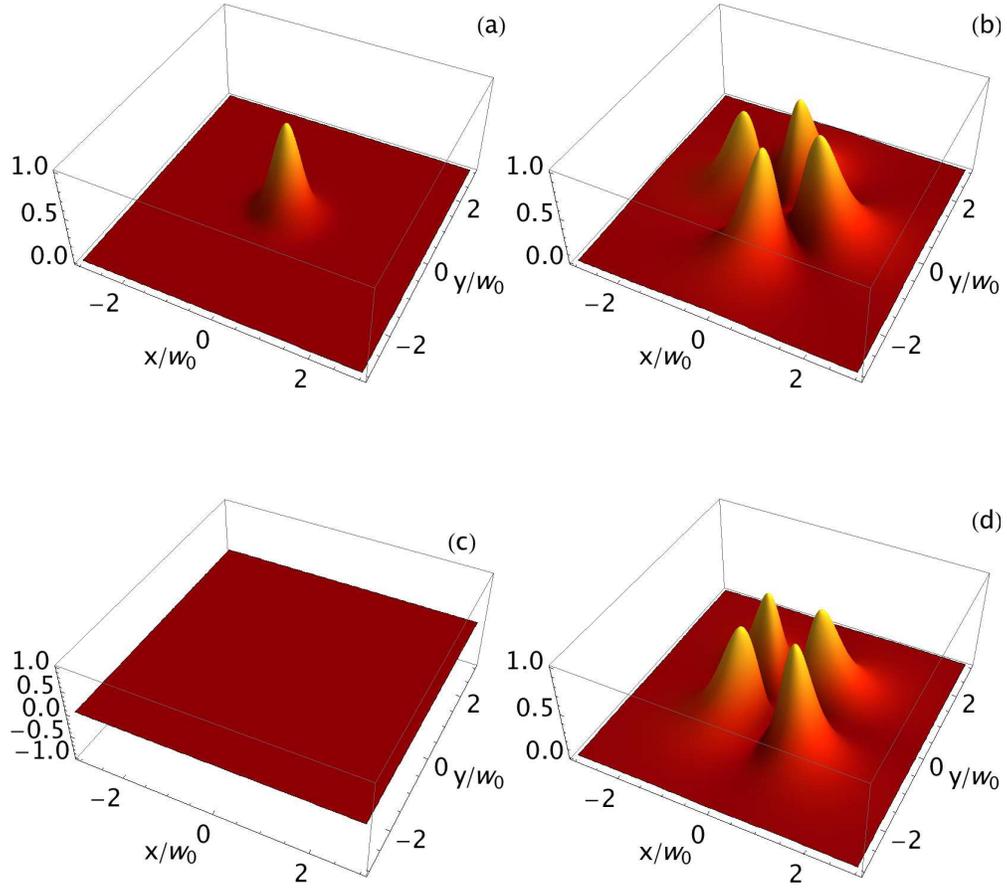}
\caption{Normalized intensity distributions for the longitudinal component $E_z(\vett{r},t)$ of the electric field of a radially (upper row) and azimuthally (lower row) polarized X-wave. Panels (a) and (c) display the co-rotating modes [Eqs. \eqref{EradP3} and \eqref{EazP3}, respectively], while panels (b) and (c) display the counter-rotating ones [Eqs. \eqref{EradM3} and \eqref{EazM3}, respectively]. These figures are plotted for $\zeta=0$, corresponding to $z=0=t$. The transverse coordinates have been normalized to the beam waist $w_0=\alpha/\sin\vartheta_0$. }
\label{figure5}
\end{center}
\end{figure}

The polarization and intensity distribution patterns for the electric field of these cylindrically polarized pulses at $\zeta=0$ are displayed in Figs. \ref{figure3} and \ref{figure4}. Moreover, due to their intrinsically nonparaxial nature, these modes admit a nonzero longitudinal component, whose intensity distribution for $\zeta=0$ is displayed in Fig. \ref{figure5} for the case of radial polarization (upper panel) and azimuthal polarization (lower panel) respectively.


\subsection{Paraxial Limit}

According to Ref. \cite{mioOptExpr} the nonparaxial modes $\Psi_{10}$ and $\Psi_{01}$ admit (in $z=0$) the Hermite-Gaussian modes $HG_{10}$ and $HG_{01}$ as paraxial limit, namely
\barr
\lim_{w_0\rightarrow\infty}\Psi_{10}(R,\theta,0; k)&=&HG_{10}(R,\theta,0; k),\\
\lim_{w_0\rightarrow\infty}\Psi_{01}(R,\theta,0; k)&=&HG_{01}(R,\theta,0; k).
\earr
Analogously, the paraxial limit of the pulses $\Phi_{10,01}(\rho,\theta,\zeta)$ in $z=0$ is also given by the correspondent Hermite-Gaussian modes. We have in fact
\barr\label{paraxialPulse}
\Phi_{10}^{(par)}(R,\theta,t)&\equiv&\lim_{w_0\rightarrow\infty}\Phi_{10}(\rho(z=0),\theta,\zeta=-ct)\nonumber\\
&=&\int_0^{\infty}\,dk\,f(k)e^{-ickt}\lim_{w_0\rightarrow\infty}\Psi_{10}(R,\theta,0; k)\nonumber\\
&=&\int_0^{\infty}\,dk\,f(k)e^{-ickt}HG_{10}(R,\theta,0; k).
\earr
A similar expression can be also obtained for $\Phi_{01}^{(par)}(R,\theta,t)$. Notice that for $z=0$, the expression of $HG_{10}(R,\theta,0; k)$ is independent on $k$, as it is given by \cite{siegman}
\beq
HG_{10}(R,\theta,0)=\frac{2\sqrt{2}R\cos\theta}{\sqrt{\pi}w_0^2}e^{-\frac{R^2}{w_0^2}}.
\eeq
We can use this result to explicitly calculate the last integral in Eq. \eqref{paraxialPulse} to obtain:
\bseq
\begin{align}
\Phi_{10}^{(par)}(R,\theta,t) &=\frac{1}{\alpha-ict}HG_{10}(R,\theta,0),\\
\Phi_{01}^{(par)}(R,\theta,t) &=\frac{1}{\alpha-ict}HG_{01}(R,\theta,0).
\end{align}
\eseq
The radially and azimuthally polarized paraxial modes can be then built following the rules derived in Sect. 3. In $z=0$, therefore, the spatiotemporal features of the X-waves become separable, and the spatial structure of the cylindrically polarized X-waves is fully equivalent to their monochromatic paraxial counterpart. For $z\neq 0$, instead, the integrals in Eq. \eqref{paraxialPulse} do not admit a closed form solution, mainly because of the complicated $k$-dependence of the Hermite-Gaussian modes. If one wants to study the propagation properties of such modes, a numerical estimation of the integral \eqref{paraxialPulse} is therefore mandatory. In alternative, one could take directly the expression of the scalar pulses $\Phi_{10,01}(\vett{r},t)$ (or, equivalently, of the correspondent vector electric and magnetic fields) and study their propagation properties as $\vartheta_0\rightarrow 0$.

\section{Conclusions}

In this work, we have generalized the method used in \cite{ragazzina} to construct cylindrically polarized modes to the domain of optical pulses. To do that, we employed nondiffracting X-waves carrying OAM. We have shown that these vector fields retain all the typical properties of their monochromatic counterpart, and analyzed their behavior both in the paraxial and nonparaxial regime. 

We believe that our work is significant both fundamental studies and for applications. In the latter case, the interplay between the nondiffracting character of X-waves and the cylindrical polarization could shine a new light on possible applications, such as material processing and particle manipulation. The fact that beams with cylindrical polarization may be tightly focalized \cite{refClaudio} may open very interesting possibilities for spatially resolved pump-probe Raman spectroscopy and related fields. On the fundamental side, as cylindrically polarized modes are often associated with classical entanglement \cite{refCe1,refCe2}, the extension of this concept to the non-monochromatic domain could open new horizons for the realization of classical and quantum hyper-entangled states.

\section*{Acknowledgements}
The authors acknowledge founding by the German Ministry of Education and Research (Center for Innovation Competence program, grant 03Z1HN31).

\end{document}